\documentclass[twocolumn,showpacs,showkeys,amsmath,amssymb,aps]{revtex4}
\usepackage{epsfig}
\usepackage{bm}

\makeatletter
\newcommand\erfc{\mathop{\operator@font erfc}\nolimits}
\def\slashchar#1{\setbox0=\hbox{$#1$}
   \dimen0=\wd0 \setbox1=\hbox{/} \dimen1=\wd1
   \ifdim\dimen0>\dimen1 \rlap{\hbox to \dimen0{\hfil/\hfil}} #1
   \else  \rlap{\hbox to \dimen1{\hfil$#1$\hfil}} / \fi}
\def\p{\slashchar{p}}

\makeatother

\begin{document}
 
\title{Landau-gauge condensates from the quark propagator on the lattice}

\author{Enrique Ruiz Arriola} \email{earriola@ugr.es}
\affiliation{Departamento de F\'{\i}sica Moderna, Universidad de
Granada, E-18071 Granada, Spain}
\author{Patrick Oswald Bowman} \email{patrick@ntc.iucf.indiana.edu}
\affiliation{Nuclear Theory Center, Indiana
University, Bloomington, Indiana 47405}
\author{Wojciech Broniowski}
\email{Wojciech.Broniowski@ifj.edu.pl} \affiliation{The
H. Niewodnicza\'nski Institute of Nuclear Physics, Polish Academy of
Sciences, PL-31342 Krak\'ow, Poland} 
\date{August-October 2004}

\begin{abstract}
We compute the dimension-2 condensate, $\langle A^2 \rangle $, and
dimension-4 mixed condensate, $\langle \bar q \slashchar{A} q \rangle $, from the recent
quenched lattice results for the quark propagator in the Landau gauge.
We fit the lattice data to the Operator Product Expansion in the ``fiducial'' 
region $1.2~{\rm GeV} \le Q \le 3~{\rm GeV}$. Our result for the dynamical
gluon mass at the scale of $10~{\rm GeV}^2$ is $ m_A=600-650~{\rm MeV}$,
in agreement with independent determinations.  For the mixed Landau gauge condensate of 
dimension-4 we get 
$\alpha_s \langle \bar q g \slashchar{A} q \rangle = (-0.11 \pm 0.03)~{\rm GeV}^4 $. This 
value is an order of magnitude larger than the $\langle G^2 \rangle$ gluon 
condensate. 
\end{abstract}

 

\maketitle 
 
\section{Introduction\label{intro}}

The question of how constituent quarks arise dynamically has always been one 
of 
the most intriguing problems of QCD.  The issue has prompted perturbative
and nonperturbative approaches both in the continuum as well as on
the lattice.  Politzer~\cite{Politzer:1976tv} was the first one to
compute the quark mass function using the Operator Product Expansion (OPE)
in the high momentum regime in terms of the quark condensate 
$ \langle \bar q q \rangle $
in the Landau gauge.  This calculation was corrected and extended to
a general Lorentz gauge by Pascual and de Rafael~\cite{Pascual:1982}. The 
gauge-independent gluon condensate $\langle G^2 \rangle$ was
included by Lavelle and Schaden~\cite{Lavelle:eg}, where it
was also forseen that a dimension-2 condensate, $\langle A^2 \rangle $,
should be present. Originally it was interpreted as a signature of spontaneous 
gauge symmetry breaking.  The dimension-4 mixed quark-gluon condensate, 
$\langle \bar q \slashchar{A} q \rangle$, was included in the
analysis of Lavelle and Oleszczuk~\cite{Lavelle:1991vr}. 

More recently, Schwinger-Dyson approaches (for reviews see,
{\em e.g.}, Refs.~\cite{Roberts:dr,Maris:2003vk,Alkofer:2000wg,Kekez:2003ri} 
and
references therein) have been intensely applied in an attempt to understand 
the 
nonperturbative physics in the infrared domain. 
The phenomenological success of
this approach has triggered a lot of activity on the lattice where
the quark propagator has recently been computed after gauge
fixing~\cite{Bowman:2002bm,Bowman:2002kn,Zhang:2003ce,Bowman:2004xi}. However,
the discussion of Ref.~\cite{Bowman:2004xi} regarding the
matching to the OPE is limited to the mass function.  Remarkably,
the dimension-2 $\langle A^2 \rangle $ condensate is related to the
quark wave function renormalization~\cite{Lavelle:eg}.

Early implications of a non-vanishing dimension-2
condensate have been explored by Celenza and Shakin 
\cite{carl1,carl2}. More recently Chetyrkin, Narison, and Zakharov  
\cite{Chetyrkin:1998yr,Narison:2001ix} found that
the inclusion of a tachyonic gluon mass parameter $ m_A \sim 700 {\rm
MeV} $ improves the phenomenology of the QCD sum rules in mesonic
channels. For heavy quarks $m_A^2 $ is proportional to the string
tension of a short string, so it provides the short-range behavior of
confining forces. Other phenomenological determinations of a
non-vanishing gluon mass can be traced from the review~\cite{Field:2001iu}. 

Although the dimension-2 $\langle A^2 \rangle $ 
condensate na\"{\i}vely breaks gauge invariance in the
perturbative sense, a detailed analysis reveals that this is not
so.  As suggested in Refs.~\cite{Gubarev:2000eu,Gubarev:2000nz,Slavnov:2004rz},
there exists a  nonlocal gauge invariant condensate,
\begin{eqnarray}
\langle A_{\rm min}^2 \rangle = \frac{1}{VT} 
	{\rm min}_g \int d^4 x \langle \left(g A_\mu g^\dagger 
	+ g \partial_\mu g^\dagger \right)^2\rangle,
\end{eqnarray} 
which reduces to the $\langle A^2 \rangle $ condensate in the Landau gauge. 
Here $g$ denotes the group element. A
physical meaning has also been attached to this condensate by a
perturbative gauge-covariant redefinition of the gluon
field~\cite{Kondo:2003uq}. Further mounting evidence for the existence
and physical relevance of the dimension-2 condensate in QCD has been also
gathered from the lattice calculations~\cite{Boucaud:2001st}, analytic
estimates~\cite{Dudal:2003vv}, purely theoretical
considerations~\cite{Gripaios:2003xq}, and microscopic
approaches~\cite{Kondo:2001nq}.  Anomalous dimensions for the $A^2$
condensate were calculated in Refs.~\cite{gr1,gr2,gr3,Boucaud:2002jt,Dudal:2002pq,ch4}.

The comparison of numerical lattice QCD calculations with analytic
continuum approaches, such as the perturbation theory, the operator product
expansion, or the Dyson-Schwinger approaches, requires a local gauge fixing
condition on the lattice. Thereafter it is possible to give a meaning
of quark and gluon correlation functions. However, it is well known
that there is no known local gauge fixing condition free of the Gribov
copies (see, {\em e.g.}, Ref.~\cite{Williams:2003du} and references
therein). Therefore, one should keep in mind that when fixing the gauge there
may still be differences in physical observables which become
non-analytic functions of the coupling constant due to the influence
of the Gribov copies. If one restricts, however, to the fundamental
modular region by a partial local gauge fixing, there may still be
gauge-invariant operators under the residual subgroup and the BRST
transformation~\cite{Gripaios:2003xq}. In the Landau gauge the only
dimension-2 operator satisfying the above condition 
is precisely $A_\mu^2$.

In the present work we extract the dimension-2 $\langle A^2 \rangle$ 
condensate by comparing
the lattice results for the quark propagator in the Landau gauge,
presented in Ref.~\cite{Bowman:2002kn,Bowman:2004xi}, to the OPE results of
Refs.~\cite{Lavelle:eg,Lavelle:1991vr}. Our determination yields a
novel estimate of the gluon mass, $m_A$, as well as provides 
the first determination of the mixed dimension-4 condensate, $ \langle \bar q g \slashchar{A}
q \rangle$ (hereafter $g$ denotes the strong coupling constant). 

\section{Lattice data for the quark propagator in the Landau gauge}

The inverse quark propagator can be parameterized as
$S^{-1}(p) = \p A(p) - B(p)$,
where $A$ and $B$, dependent on the quark momentum, have the meaning of the vector and scalar quark
self-energies.  An equivalent parameterization is via the
wave-function renormalization, $Z$, and the mass function, $M$,
defined as
\begin{eqnarray}
S(p)&=&\frac{Z(p)}{\p - M(p)}, \label{MZpar1} \\
Z(p)&=&A^{-1}(p), \;\;\; M(p)=B(p)/A(p).
\label{MZpar}
\end{eqnarray}

The quark propagator was calculated in Landau gauge using the ``Asqtad'' 
improved staggered action.  The gauge ensemble is made of 100 quenched, 
$16^3\times 32$ lattices with a nominal lattice spacing $a = 0.124~\text{fm}$, 
set from the static quark potential.  This data was first published in 
Ref.~\cite{Bowman:2002kn}.

\begin{figure}[]
\begin{center}
\epsfig{figure=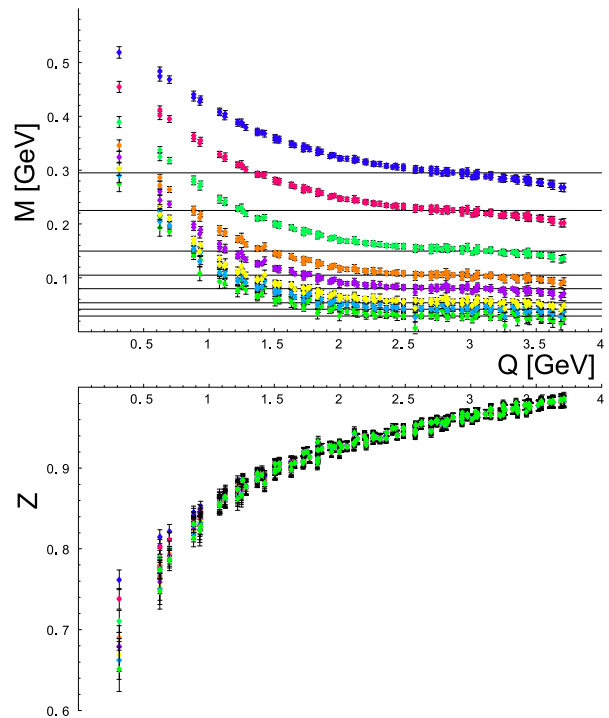,width=8cm}
\end{center}
\vspace{-5mm}
\caption{The quark mass function $M$ (top) and the wave-function
renormalization $Z$ (bottom), plotted as functions of the Euclidean
momentum $Q$. The data comes from quenched lattice calculations
in the Landau gauge of Ref.~\cite{Bowman:2002bm}. Various sets of
points correspond to the current quark masses $m=29$, $42$, $54$,
$80$, $105$, $150$, $225$, and $295$~MeV, indicated by horizontal
lines in the top panel.  In both panels the highest sets of points
correspond to highest values of $m$.}
\label{fig:MZ}
\vspace{-5mm}
\end{figure}

The results for $M$ and $Z$ as functions of the Euclidean momentum $Q$
are shown in Fig.~\ref{fig:MZ} at various values of the current quark
mass $m$.  The data for $M$ asymptote at large $Q$ to the value of
$m$, indicated by the horizontal lines.  We note that the data at
highest values of $Q$ are not perfect, with some visible wiggles and a
tendency of falling off at the end, which may be attributed to the
finite-size effect.  Yet, up to $Q\sim 3~{\rm GeV}$ the tails in $M$
and $Z$ look very reasonably, reaching plateaus before ``hitting the
wall''.

As a matter of fact, the tail in $M$ in the ``fiducial'' region of
$1.9~{\rm GeV} \le Q \le 2.9~{\rm GeV}$ was used successfully in
Ref.~\cite{Bowman:2004xi} to verify the expression
\begin{eqnarray}
M(Q)&=&-\frac{4\pi^2 d_M \langle \bar q q \rangle_\mu
[\log(Q^2/\Lambda_{\rm QCD}^2)]^{d_M-1}} {3 Q^2
[\log(\mu^2/\Lambda_{\rm QCD}^2)]^{d_M}} \nonumber \\ &+&
\frac{m(\mu^2)[\log(\mu^2/\Lambda_{\rm QCD}^2)]^{d_M}}
{[\log(Q^2/\Lambda_{\rm QCD}^2)]^{d_M}},
\label{mas}
\end{eqnarray} 
where $d_M=12/(33-2N_f)$ with $N_f=0$ flavors, $\langle \bar q q \rangle_\mu$
and $m(\mu)$ are the quark condensate and the current quark mass at
the scale $\mu$, respectively, and $\Lambda_{\rm QCD}=691$~MeV in the
$\text{MOM}$ scheme. This shows that the data is accurate enough
to be verified against the perturbative QCD predictions.

The data for $Z(Q)$ from Ref.~\cite{Bowman:2002bm} are shown in the bottom
panel of Fig.~\ref{fig:MZ}.  A very weak dependence on $m$ has been noted,
except perhaps at low $Q$. Asymptotically, $Z(Q) \to 1$, as requested
by the canonical normalization of the quark fields. At lower values of
$Q$ the departure of $Z$ from unity is sizeable, with a long-range
tail clearly visible.

\section{Matching OPE to lattice data \label{sec:chi2}}

\begin{figure}[]
\begin{center}
\epsfig{figure=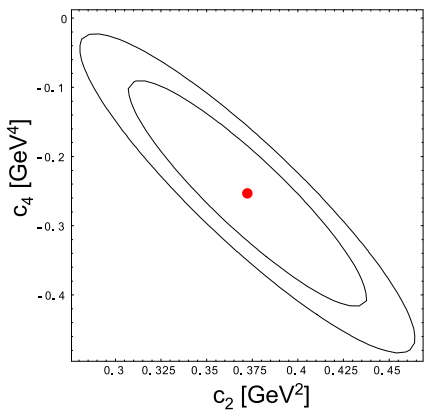,width=5.7cm}
\end{center}
\vspace{-7mm}
\caption{The $\chi^2$ contours corresponding to the fit of the data of 
Ref.~\cite{Bowman:2002bm}  for the vector self-energy $A(Q)$ in the Landau gauge
to formula~(\ref{aexp}). The dot indicates the optimum 
values of Eq.~(\ref{fit}). The curves correspond to 68\% and 95\% 
confidence levels. Note a large correlation between $c_2$ and $c_4$.}
\label{fig:chi2}
\end{figure}

In our further analysis we will work with the function $A(Q)$.
The data for the vector quark self-energy $A(Q)$ may be parameterized at
sufficiently large values of $Q$ as
\begin{eqnarray}
A(Q)=1+\frac{c_2}{Q^2}+\frac{c_4}{Q^4}.
\label{aexp}
\end{eqnarray} 
\begin{figure}[]
\begin{center}
\epsfig{figure=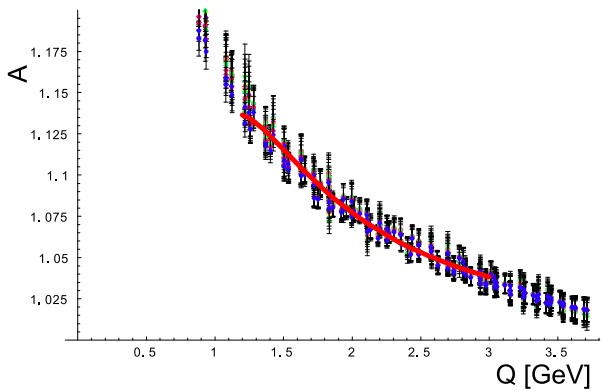,width=8cm}
\end{center}
\vspace{-7mm}
\caption{The fit to the tail of the vector quark self-energy, $A(Q)$.
The solid line corresponds to the asymptotic formula (\ref{aexp}) with
the optimum parameters (\ref{fit}), while the data (including all
values of the current quark masses $m$) are taken from the 
quenched lattice calculation in the Landau gauge of
Ref.~\cite{Bowman:2002bm}.  The asymptotic curve is drawn in the fiducial region
of $1.2~{\rm GeV} \le Q \le 3< {\rm GeV}$.}
\label{fig:Afit}
\end{figure}
In the fitting procedure we must decide on the matching region in $Q$. Certainly,
this choice will affect the results, yielding a systematic error. The values of 
$Q$ cannot be too large due to finite-size effects, nor too small, where the 
expansion~(\ref{aexp}) is no longer accurate. We perform the $\chi^2$ fit in the 
range $1.2~{\rm GeV} \le Q \le 3~{\rm GeV}$, which yields the optimum values 
\begin{eqnarray}
c_2 = (0.37 \pm 0.04)~\text{GeV}^2, \;
c_4 = (-0.25 \pm 0.06)~\text{GeV}^4.
\label{fit}
\end{eqnarray}
The errors have been calculated by jackknife.  The value of $\chi^2/{\rm DOF}$ 
is $0.51$, but one can see from the plot of 
Fig.~\ref{fig:chi2} that there is a sizeable correlation between $c_2$ and $c_4$. 

We have also performed the fit with three terms, appending formula
(\ref{aexp}) with the term $c_6/Q^6$. The result is $c_2=0.39~{\rm
GeV}^2$, $c_4=-0.37~{\rm GeV}^4$, and $c_6=0.15~{\rm GeV}^6$ with errors that
overlap with the central values for $c_2$ and $c_4$ of Eq.~(\ref{fit}).
However, due to large correlations between $c_4$ and $c_6$, no reliable
information may be extracted from this three-parameter fit.
More accurate data and a larger range of momenta will allow for a better 
determination of the $1/Q^2$-expansion.

Next, we will compare the obtained values of Eq.~(\ref{fit}) to
theoretical predictions and extract estimates for the Landau-gauge 
condensates.  At $D=4$ the vector self-energy read out
from the propagator of Ref.~\cite{Lavelle:eg,Lavelle:1991vr} is
\begin{eqnarray} 
A(Q)&=&1+\frac{\pi \alpha_s(\mu^2) \langle A^2 \rangle_\mu}{N_c Q^2}
\\ &-& \frac{\pi \alpha_s(\mu^2) \langle G^2 \rangle_\mu}{3 N_c Q^4} +
\frac{3\pi \alpha_s(\mu^2) \langle \bar q g \slashchar{A} q
\rangle_\mu}{4 Q^4}, \nonumber
\end{eqnarray}
where $\mu$ denotes the renormalization scale.
Comparing to Eq.~(\ref{fit}) we find for three colors
\begin{eqnarray}
\alpha_s(\mu^2) \langle A^2 \rangle_\mu = (0.36 \pm 0.04)~{\rm GeV}^2, \label{condA2}
\end{eqnarray}
or
\begin{equation}
g^2 \langle A^2 \rangle = (2.1 \pm 0.1 ~{\rm GeV})^2, \label{gcondA2}
\end{equation}
and
\begin{eqnarray}
\alpha_s(\mu^2) \langle \bar q g \slashchar{A} q \rangle_\mu - 
\frac{4\pi}{27} \langle \frac{\alpha_s}{\pi}G^2\rangle 
   = (-0.11 \pm 0.03)~\text{GeV}^4. \nonumber \\
\label{condA4}
\end{eqnarray}
Since  
$\langle \frac{\alpha_s}{\pi} G^2 \rangle \simeq 0.009~{\rm GeV}^4$
\cite{Ioffe:2002be}, the contribution of the
$\langle G^2 \rangle$ condensate to Eq.~(\ref{condA4}) is negligible
compared to the mixed-condensate term. Thus
\mbox{$\alpha_s(\mu^2) \langle \bar q g \slashchar{A} q \rangle_\mu =(-0.11
\pm 0.03)~{\rm GeV}^4$}.

The errors quoted throughout the paper are statistical. In addition, there are
certain systematic errors originating from the choice of the fitted function 
$A(Q)$ of Eq.~(\ref{aexp}) and from the choice of the fiducial region in $Q$. 
Quantities quoted in physical units are also subject to the uncertainty in 
scale that goes with quenched lattice simulations.

\section{Comparison of $\langle A^2 \rangle$ to other determinations \label{sec:ev}} 

The Landau-gauge condensates considered in this paper are not
renorm-invariant quantities, thus their values evolve perturbatively
with the scale.
The QCD evolution for $\langle A^2 \rangle$ has been worked out in
Ref.~\cite{gr1,gr2,gr3,Boucaud:2002jt,ch4}, with the leading-order result
\begin{eqnarray}
\alpha_s(\mu^2) \langle A^2 \rangle_\mu \sim
\alpha_s(\mu^2)^{1-\gamma_{A^2}/\beta_0}, \label{evol}
\end{eqnarray}
where $\gamma_{A^2}=35/4$ and $\beta_0=11$ correspond to evolution
with no flavor. We use $\alpha_s(\mu^2)=4 \pi/(9 \log [ \mu^2/\Lambda^2])$, 
with $\Lambda=226~{\rm MeV}$ for the evolution at the leading order.
The exponent in Eq.~(\ref{evol}) is equal to 9/44,
hence the evolution is very slow. For instance, the change of $\mu^2$
from $1~{\rm GeV}^2$ up to $10~{\rm GeV}^2$ results in a reduction of
$\alpha_s \langle A^2 \rangle$ by 10\% only.

Most estimates in the literature refer to the gluon mass, related to 
the $\langle A^2 \rangle$ by the formula
$m_A^2=\frac{3}{32}g^2 \langle A^2 \rangle$.
Our estimate (\ref{condA2}), when evolved with Eq.~(\ref{evol}) from 2 to 10~GeV$^2$, yields
\begin{eqnarray}
m_A=(625 \pm 33)~{\rm MeV}. \label{mg}
\end{eqnarray}
Evolution from 1 to 10~GeV$^2$ gives $m_A=(611 \pm 32)~{\rm MeV}$, while 
evolution from 4 to 10~GeV$^2$ produces $m_A=(635 \pm 34)~{\rm MeV}$. 
These values are close to many estimates made in other approaches. 
In particular, most of the numbers listed in Table~15 of the review of 
Ref.~\cite{Field:2001iu} and obtained by very different techniques
are in the range 0.5-1.5~GeV. 

\section{Conclusions}

We have attempted to match the OPE to the quenched 
lattice data for the vector quark energy in the Landau gauge.
The obtained value of the dimension-2 Landau-gauge gluon condensate, 
$\langle A^2 \rangle$, of Eq.~(\ref{condA2}) and the corresponding estimate for
the gluon mass of Eq.~(\ref{mg}) are consistent with other estimates 
in the literature.  Thus the lattice provides an independent way of 
determining this condensate.
The estimate for the dimension-4 mixed quark-gluon condensate of 
Eq.~(\ref{condA4}), made to our knowledge for the first time, is an order of 
magnitude larger compared to the $\langle G^2 \rangle$ condensate.

\begin{acknowledgments}
Support from DGI and FEDER funds, under contract BFM2002-03218 and by the
Junta de Andaluc\'\i a grant no. FM-225 and EURIDICE grant number
HPRN-CT-2003-00311 is acknowledged.  Partial support from the Spanish
Ministerio de Asuntos Exteriores and the Polish State Committee for
Scientific Research, grant number 07/2001-2002 is also gratefully
acknowledged.
\end{acknowledgments}

\end{document}